# Mechanical Properties and Nanostructure of Monolithic Zeolitic Imidazolate Frameworks: A Nanoindentation, Nanospectroscopy and Finite-Element Study

*Michele Tricarico* and *Jin-Chong Tan*[*]

*Multifunctional Materials and Composites (MMC) Laboratory,*
*Department of Engineering Science, University of Oxford, Parks Road,*
*Oxford, OX1 3PJ, United Kingdom.*
[*]E-mail: jin-chong.tan@eng.ox.ac.uk

## Abstract

The synthesis of metal-organic frameworks (MOFs) in a monolithic morphology is a promising way to achieve the transition of this class of materials from academia to industrial applications. The sol-gel process has been widely employed to produce MOF monoliths. It is relatively cheap and simple compared to other techniques (*e.g.,* mechanical densification) and moreover it allows to produce "pure" monoliths, *i.e.,* without the need of using binders or templates that could affect the functional properties of the MOF. Understanding the mechanical properties of these monoliths is crucial for their transit to practical applications. We studied the mechanical behavior of two zeolitic imidazolate frameworks (ZIF-8 and ZIF-71) by means of instrumented nanoindentation and atomic force microscopy (AFM). Tip Force Microscopy (TFM), an extension of AFM, was used to reveal the surface nanostructure of the monoliths. We employed finite-element (FE) simulations alongside the experiments, to establish a suitable constitutive model and determine an improved estimate of the yield stress ($\sigma_Y$) of ZIF monoliths. NanoFTIR was subsequently used to pinpoint local structural alteration of the framework in the contact area. The combination of TFM, FE simulations, and nanoFTIR enabled us to identify the mechanical deformation mechanisms in monolithic ZIF materials: grain boundaries sliding is dominating at low stresses, then breakage of chemical bonds and a partial failure of the framework occurs, eventually leading to a densification of porous framework at the contact zone. Finally, we measured the fracture toughness using a cube corner indenter to study the resistance of monoliths against cracking failure.

Metal-organic frameworks (MOFs) are an emerging class of hybrid porous materials. They consist of metal clusters and organic ligands that self-assemble to form a lattice framework with a considerable internal surface area. Research over the last 25 years has led to a vast range of framework-type compounds with high chemical stability, tunable physical properties, organic functionality, and porosity,[1, 2] resulting in several possible technological applications.[3-6]

The transition of these materials from academic research to industrial practical applications is however rather limited. One of the reasons of this deficiency is a limited understanding of the mechanical properties. Despite the large amount of work devoted to the chemical synthesis of novel framework structures and characterization of their functional properties, the field of MOF mechanics is still at an early stage.[7-9]

The mechanical characterization of MOFs has mostly relied on nanoindentation measurements, which allow for the possibility of testing very small samples, such as single crystals,[10-14] thin films,[15] or small monoliths.[16, 17] In particular, Zeolitic Imidazolate Frameworks (ZIFs)[18] have been extensively studied by nanoindentation, due to their ease of synthesis, stability under ambient conditions and commercial availability of certain compounds.[11, 12, 14]

The majority of MOFs are synthesized as polydisperse microcrystalline powders. These may present intrinsic limitations, such as poor handling properties, restricted mass transfer, and mechanical instability, resulting in a further narrowing of the range of industrial applications. The realization of mechanically robust and bulk morphologies, known as "monoliths", is therefore the preferred form factor for commercial applications.[19, 20] The advantages of monolithic structures are multiple: easier handling associated with structural rigidity, low mass transfer resistance (for improved gas separation capabilities) and higher volumetric adsorption capacities.[16] MOF monoliths have been fabricated using different techniques, such as coating of organic[21] and inorganic[22] scaffolds, extrusion by means of binders and plasticizers,[23] 3D printing,[24, 25] mechanical densification,[26-28] template morphological replacement [29] and sol-gel processes.[19] With this technique, it is possible to tune the hierarchical pore distribution within the monolith, leading to improved capabilities for gas uptake.[16, 17, 20, 30]

Several classes of MOFs have already been studied in their sol-gel monolithic form, including HKUST-1,[16] ZIFs[17, 31] and UiO-66.[30] The sol-gel process consists of the production of a network (gel) from molecular precursor *via* the formation of colloidal particles (sol). The nanoparticles in the colloid aggregates *via* weak non-covalent interactions (mainly *van der Waals* forces), forming a viscous material: the gel. Upon drying, the gel progressively transforms into a monolith, as illustrated in the scheme in *Figure 1a*. The drying process plays undoubtedly a prominent role on the formation of mechanically stable monoliths. During solvent drying, there exists a mechanical stress at the gas-liquid interface in the pores, due to surface tension. This phenomenon is often the cause of cracking or collapse of the gel body.[32] A slow solvent removal (typically at room temperature) and small particles size can better accommodate this stress, promoting the formation of dense monoliths. However, this is not always possible, especially with big monoliths, where large shrinkages occur, leading to cracking anyway. Another crucial requirement for the formation of a gel rather than a precipitate, is a small size of the particles. This can be achieved during the crystallization phase, by increasing the number of nucleation sites. As a result, the crystal growth will be constrained, leading to the formation of fine particles (usually smaller than 100 nm).[19]

In this work, we aim to study the detailed mechanical behavior of the ZIF-8 and ZIF-71 sol-gel monoliths. To achieve the gel state, the HCR (high concentration reaction) synthetic approach, recently developed by our group[33, 34] was leveraged. This method consists of the accelerated deprotonation of the organic linker, which boosts the formation of metal-organic clusters and hence increases the number of nucleation sites, leading to the formation of fine nanoparticle size that promotes the formation of the gel. Upon drying of the gel, these particles aggregate and consolidate to form the monolith. Tip Force Microscopy (TFM) imaging of the surface of the monoliths revealed a fine-grained microstructure (grain size < 100 nm) or "nanostructure", as shown in *Figures 1b* and *1c*, which gives us significant information to establish their structure-mechanical property relationships.

Moreover, we estimated the yield stress ($\sigma_Y$) of a monolithic MOF material, which underpins the elastic-plastic transition. To do so, we performed Finite Element (FE) simulations alongside the nanoindentation experiments to gain a better understanding of the underpinning mechanics. To the best of our knowledge, there is not yet any experimental/modelling study proposing a plasticity model for MOF monoliths and an estimate of the yield stress ($\sigma_Y$), which is a key material parameter in structural design of load-bearing applications. The nanostructure of the monoliths was mapped by TFM, which, together with the results of the FE simulations and the nanoFTIR measurement of the local regions of residual indents, allowed us to propose the deformation mechanisms of ZIF monoliths under stress.

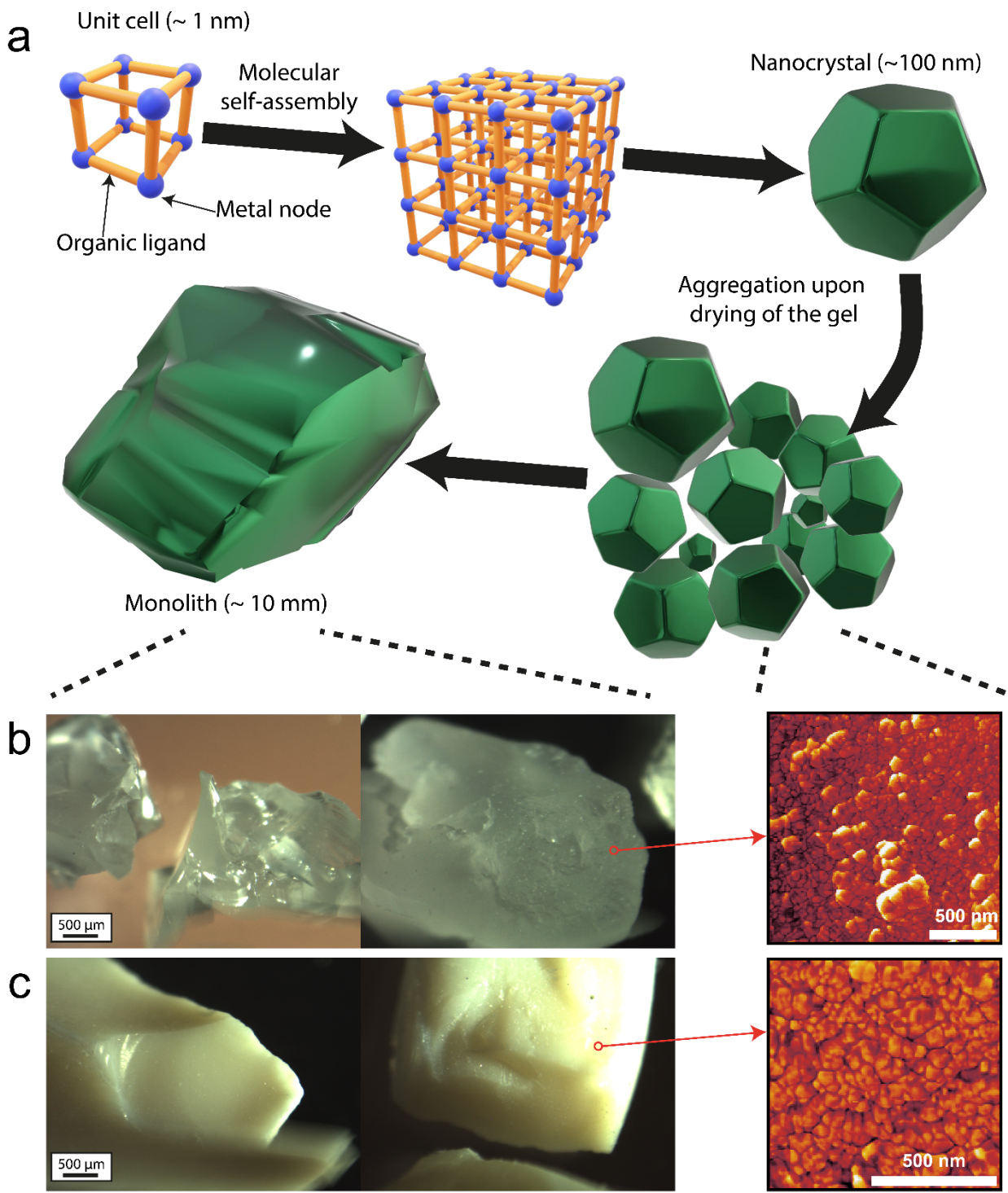

***Figure 1.*** *(a) Schematics of the formation of the MOF monoliths. (b) and (c): Optical images of ZIF-8 and ZIF-71 monoliths, respectively. Right panels show the stiffness maps of the surface of the monoliths obtained by TFM, where the nanosized grain structure is prevalent.*

**Results and discussion**

*Nanoindentation experiments and finite element simulations*

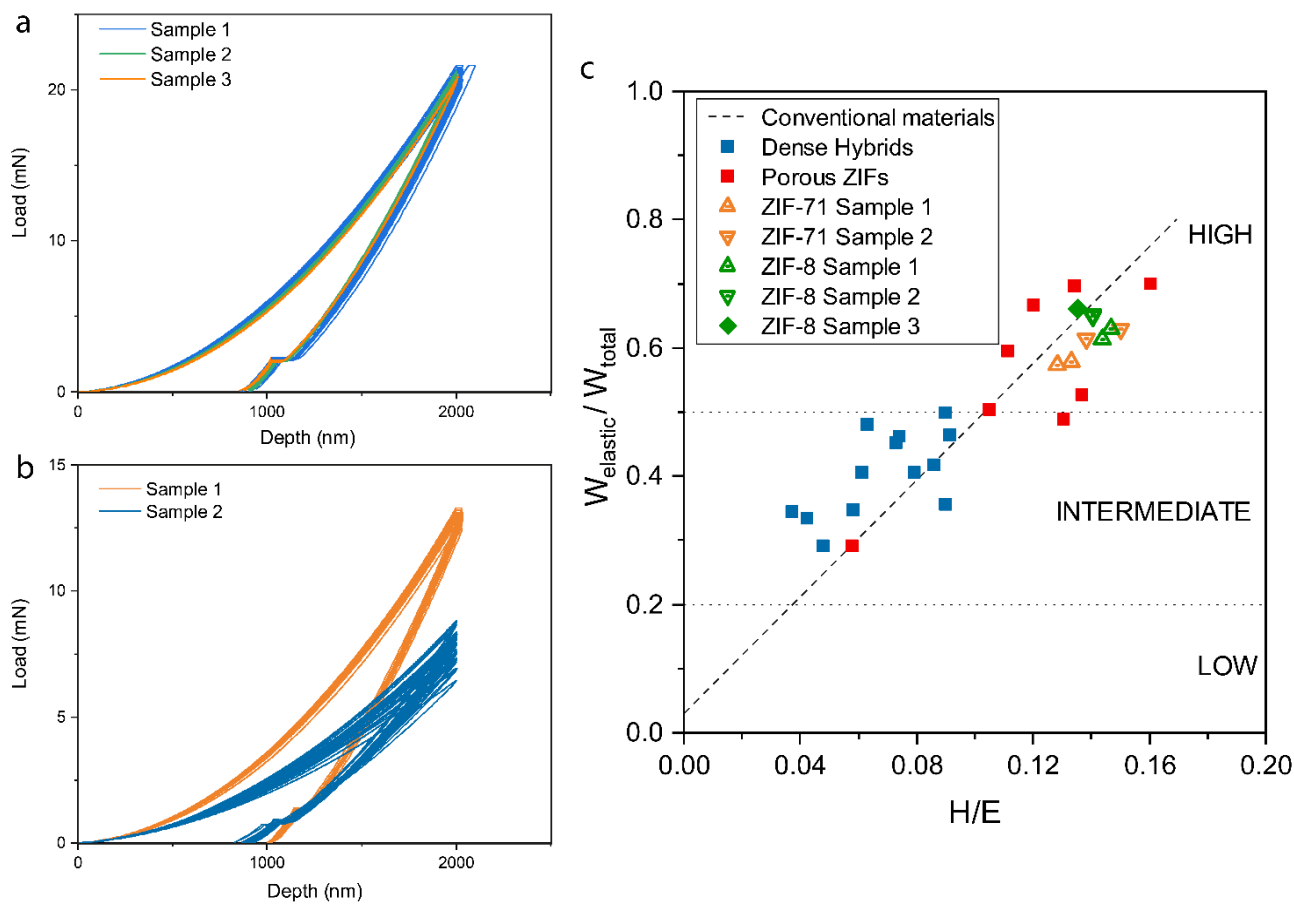

*Figure 2.* Load-depth curves resulting from nanoindentation tests of (a) ZIF-8 and (b) ZIF-71 with a Berkovich indenter and a map of elastic recovery vs the ratio of hardness to Young's modulus (H/E) (c). The data for conventional materials, dense hybrids and porous ZIFs were adapted from reference [13].

The results of a series of systematic nanoindentation measurements on the ZIF-8 monoliths are summarized in *Figure 2*. The load-depth curves resulting from nanoindentation of ZIF-8 monoliths with the Berkovich tip are shown in *Figure 2a*. A total of 59 tests on three different samples are considered. The load-depth curves show a very good repeatability, as well as the resultant values of Young's modulus ($E$) and hardness ($H$) with a relatively small standard deviation. Assuming a Poisson's ratio $v = 0.4$, the Continuous Stiffness Measurements (CSM) method gave the results listed in *Table 1*. The measured values of $E$ and $H$ are consistent with those observed before, for either single crystals[7, 11] or monoliths.[17] In order to get some insights about the mechanical behavior of the monoliths, we compare the measurements with the literature available regarding the single crystal form of the same frameworks. Interestingly, we found that the Young's modulus of the ZIF-8 monolith is reminiscent of the single crystal counterpart. This means that the nanocrystals forming the monoliths are efficiently packed and intergranular porosity is limited. The reduced hardness of the monolith (declined by ~15% compared with single crystal) is justified by the presence of "grain" boundaries between the adjacent nanocrystals forming the monolith, where the plastic deformation is more likely to initiate from a grain boundary "defect". The good repeatability of the tests obtained from the representative samples allows us to confidently classify the ZIF-8 monolith as mechanically isotropic.

In contrast, the nanoindentation of ZIF-71 monoliths resulted in a wider spread of load-depth curves (*Figure 2b*) for each sample, indicating that this monolith is less homogeneous and non-isotropic. The resultant values of *E* and *H* (*Table 1*) are appreciably lower than those of ZIF-8 monoliths. To gain further insights, we measured the density of the two monoliths (*via* Archimedes' principle, see Methods). We found the ZIF-8 monolith to be denser than ZIF-71 (1.260 vs. 0.975 g/cm$^3$) as expected, since the density of porous solids scales with its Young's modulus ($E \propto \rho^2$).[12] As reported by *Tian et al.*,[17] the density of ZIF-8 monolith is higher than the calculated theoretical density of the ideal framework structure (*i.e.* 0.95 g/cm$^3$),[12] due to presence of unreacted species and impurities. Interestingly, the ZIF-71 monolith, as opposed to ZIF-8, shows a lower density compared to the theoretical value of 1.16 g/cm$^3$ (calculated from crystal structure, CCDC code GITVIP). This observation, consistently with the nanoindentation results, gives evidence of the more efficient packing of the ZIF-8 nanocrystals compared with the ZIF-71 counterpart. In fact, the poorer packing of ZIF-71 nanocrystals in monolith results in intergranular porosity, which is evident from TFM images presented in the following section.

| Sample | # of indents | Young's modulus, *E* (GPa) | Hardness, *H* (MPa) |
|---|---|---|---|
| ZIF-8 monolith | 59 | 3.18 ± 0.04 | 452 ± 20 |
| ZIF-8 single crystal, Ref.[12] | - | 3.199 ± 0.092 | 531 ± 28 |
| ZIF-71 monolith (total) | 50 | 1.67 ± 0.38 | 227 ± 47 |
| ZIF-71 sample #1 | 25 | 2.04 ± 0.06 | 273 ± 7 |
| ZIF-71 sample #2 | 25 | 1.29 ± 0.07 | 180 ± 5 |

*Table 1. Young's modulus and hardness measurements (CSM method) obtained via Berkovich nanoindentation. Mean values and standard deviations from this work were calculated from an indentation depth range of 500-2000 nm.*

From the load-depth curves, we also computed the elastic recovery, $W_e$, defined as the ratio between the area under the unloading and loading curves respectively:

$$W_e = \frac{W_{\text{elastic}}}{W_{\text{total}}} = \frac{\int_{h_f}^{h_{\max}} P_{\text{unloading}} \, dh}{\int_0^{h_{\max}} P_{\text{loading}} \, dh} \quad (1)$$

where $h_{\max}$ is the maximum indentation depth, $h_f$ the residual depth after unloading, and $P_{\text{unloading}}$ and $P_{\text{loading}}$ are the loads applied upon loading and unloading, respectively. We obtained the value $W_e$ = 64.1 ± 1.8 % for ZIF-8 monoliths and $W_e$ = 59.9 ± 2.4 % for ZIF-71 monoliths, which are consistent with values previously reported for porous ZIFs.[13] According to the classification proposed by Coates et al.,[13] our monoliths follow the trend observed for other conventional materials[35] and fall into the category of materials with a high elastic recovery (*Figure 2c*). This result, together with the Young's modulus, suggests that the elastic response of the framework is not affected by the granular structure of the monoliths.

FEM simulations of the indentation test of ZIF-8 monoliths were performed, by modelling the indenter as a rigid 2D axisymmetric cone (*Figure 3b*); see *Methods* for further details. Since the monolithic sample was modelled as a homogeneous isotropic "substrate", we calculated Young's modulus and hardness using the standard Oliver and Pharr method,[36] instead of using the averaged value from CSM, which takes into account local effects. *Equations 3* and *5* in *Methods* were employed to calculate the modulus, where the contact stiffness ($S$) is the derivative of the unloading curve at maximum load. The Poisson's ratio ($v$) was set to 0.4, consistent with the experiments. The hardness was computed by *Equation 4*. From FE modelling of nanoindentation, we determined that $E$ = 2.87 GPa and $H$ = 420 MPa for the ZIF-8 monolith, and $E$ = 1.80 GPa and $H$ = 270 MPa for the ZIF-71 monolith.

According to *Equation 6*, the yield stress is directly related to the hardness *via* the constraint factor ($C$), which is not known *a priori* for our material, therefore we need to make an assumption. We attempted different values of $C$ and iterate to seek for the best fit as the actual value of $C$, which usually lies in the range 1.5 to 3 for conventional materials. We found the best fit to be $C$ = 2.1 (*Figure S4*), resulting in $\sigma_Y$ = 200 MPa for ZIF-8 and $\sigma_Y$ = 130 MPa for ZIF-71. Subsequently, the input material parameters for the simulations were set to be $E$ = 2.87 GPa, $v$ = 0.4 and $\sigma_Y$ = 200 MPa for ZIF-8. For ZIF-71, the material parameters were set to be $E$ = 1.80 GPa, $v$ = 0.4 and $\sigma_Y$ = 130 MPa.

The monolithic sample was modelled with two different material constitutive relations: the elastic-perfectly plastic model and the Drucker-Prager model. The latter model is suited for pressure-dependent materials, whose yield behavior depends on the hydrostatic pressure. The elastic-perfectly plastic model matches very well the experimental data in the case of the Berkovich and spherical indenters for both ZIF-8 and ZIF-71. In the case of cube corner, which is sharper than the Berkovich indenter, it can be seen in *Figure 3a* that such a model notably underestimates the maximum load. This can be explained by the effect of densification, which causes a reduction of the volume of material in the region immediately underneath the indenter under the effect of the hydrostatic pressure. Structural compression may occur due to the collapse of nanopores in the framework. We established that this effect is significant only in the case of indentation with a cube corner indenter, which is characterized by a larger hydrostatic pressure generated under the indenter. According to Tan *et al.*,[11] the amorphization of the ZIF-8 framework can be triggered by a shear stress, given its exceptionally low shear modulus, $G_{minimum} \lesssim 1$ GPa. We compared the shear stress fields at the maximum indentation depth (2,000 nm) under the Berkovich, cube corner, and spherical indenters (*Figures 3d, e, f* respectively). Similar contour plots were extracted for simulated ZIF-71 monoliths, showing the same trend but values scaled down in proportion to $\sigma_Y$ (*Figure S5*). The shear stresses predicted immediately under the cube corner indenter is higher than the other two, meaning that ZIF-8 has the greater propensity to lose its crystallinity in this case. We propose that the framework densification is initiated by the shear stress that excessively distorts the porous framework (when shear yield stress, $\tau_{max} \gtrsim \sigma_Y/2$) triggering structural amorphization, which is then susceptible to undergo densification under the influence of a hydrostatic pressure.

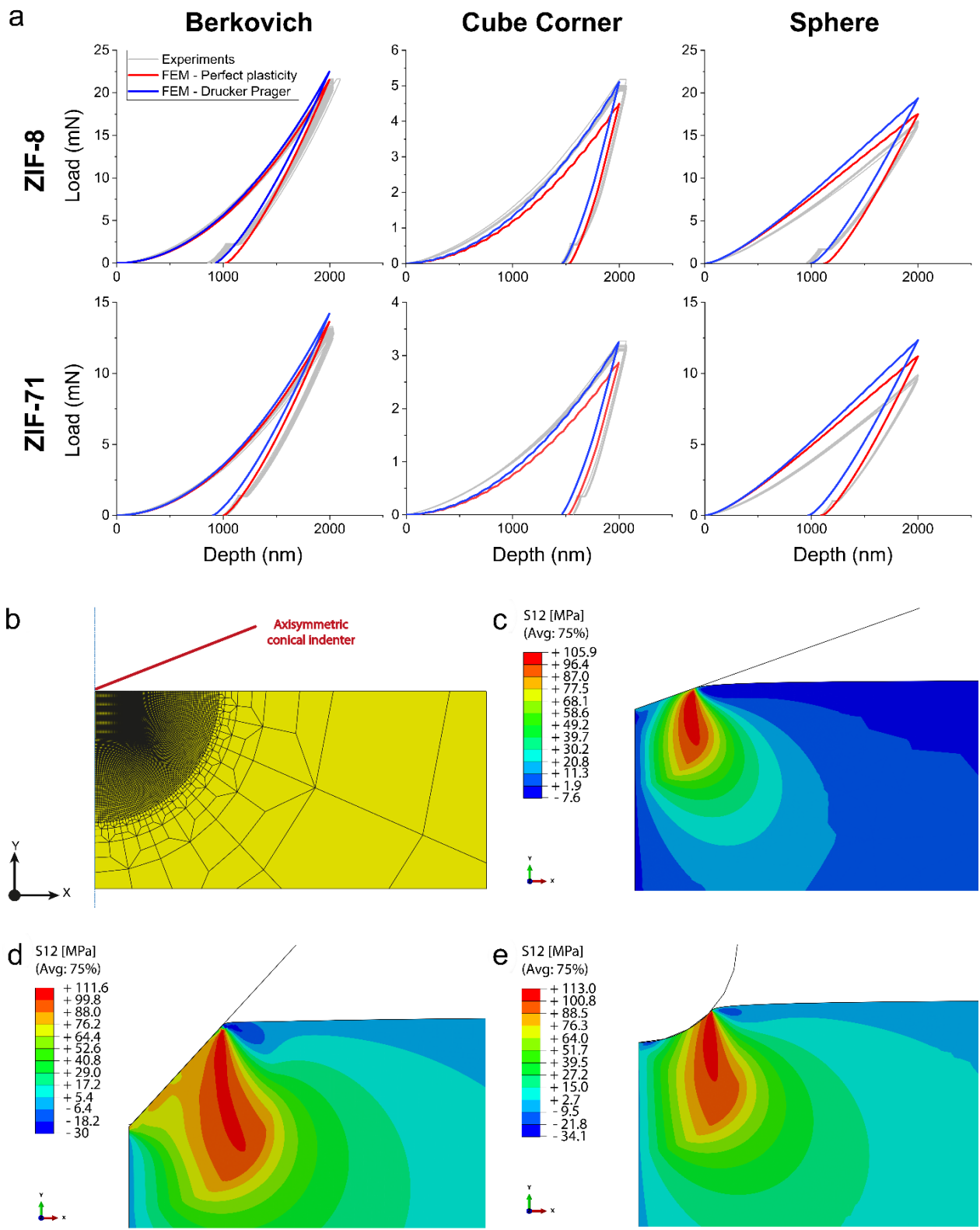

***Figure 3.*** *Finite element modelling of the nanoindentation tests. (a) Predicted nanoindentation load-depth curves of the 2-D FEM model compared to the experimental data. The material behavior was simulated with two different material models (elastic-perfectly plastic and Drucker-Prager). (b) 2-D axisymmetric FE model: the Berkovich indenter was modelled as a semi-apical cone of 70.3° (red), the sample (yellow) was meshed using a finer element size in the proximity of the contact. (c-e) Shear*

*stress contours of ZIF-8 monolith model at a maximum indentation depth (2000 nm) under the Berkovich, cube corner, and spherical indenters, respectively. For ZIF-8, the shear yield stress $\tau_{max}$ is estimated as $\sigma_Y/2$, such that the contour regions ≳ 100 MPa are predicted to have yielded under shear. The Berkovich and cube corner indenters were modelled as equivalent 2-D cones with semi-apical angles of 70.3° and 42.3°, respectively (see Methods).*

### Characterization of nanograined structures

TFM was used to image the surface of the monoliths. A nano-grained microstructure, which we termed "nanostructure" was observed for the monoliths, see *Figure 4*. Each nano-grain is in fact a single crystal, and its size is below 100 nm. This reduced grain size suggests that the main deformation mechanism is grain boundary sliding (GBS), at least until a threshold value of stress, when, according to FE simulations, a certain amount of densification occurs. This mechanism is confirmed by nanoFTIR spectra taken across the residual indents. As it is visible from *Figure 5*, the main ZIF-8 peaks remain unchanged in the Berkovich indent, but in the center, where the pressure is expected to be the highest, the infrared vibrational peaks at ~1311 cm$^{-1}$ become more intense. According to a recent study,[37] this particular peak can be associated with structural defects, namely the missing imidazolate-type linker, induced by high stress or pressure. This phenomenon is even more pronounced inside the cube corner indent, where the stress level reached upon loading is the highest, thereby resulting in bond breakage as evidenced by the local broadening of the nanoFTIR spectra shown in *Figure 5e*.

AFM was employed to observe the geometry of the residual pile-ups. As illustrated in *Figure 6*, the residual indents left by each of the indenters in both ZIF-8 and ZIF-71 monoliths, exhibit quite an insignificant pile-up. A small amount of pile-up is usually explained with a negligible work hardening, here facilitated by GBS mechanism, confirming that the elastic-perfectly plastic model represents a satisfactory approximation of the constitutive behavior of these monoliths. From a microstructural point of view, the absence of work hardening can be explained by to the small size of the nanograins. Such a nanostructure also explains the absence of cracking in the vicinity of the indents. The reduced size of the grains facilitates the GBS mechanism. A similar behavior is observed for some conventional nanocrystalline material, *e.g.* nanocrystalline ceramics, that exhibit ductility - perhaps even superplasticity - in compression, while their larger grained counterparts are extremely brittle.[38] The ductility may derive from the grain boundary sliding and rotation, producing a flow of material.[39] For example, Karch *et al.*[40] indented (with a cube corner indenter) two samples of TiO$_2$, a conventional polycrystal and a nanocrystalline one: the first showed a brittle behavior, characterized by cracking and high hardness, while the latter exhibited a lower hardness and a ductile behavior, evidenced by the absence of cracking. We propose that this type of nanograined-induced phenomenon can be extended to our monolithic ZIFs.

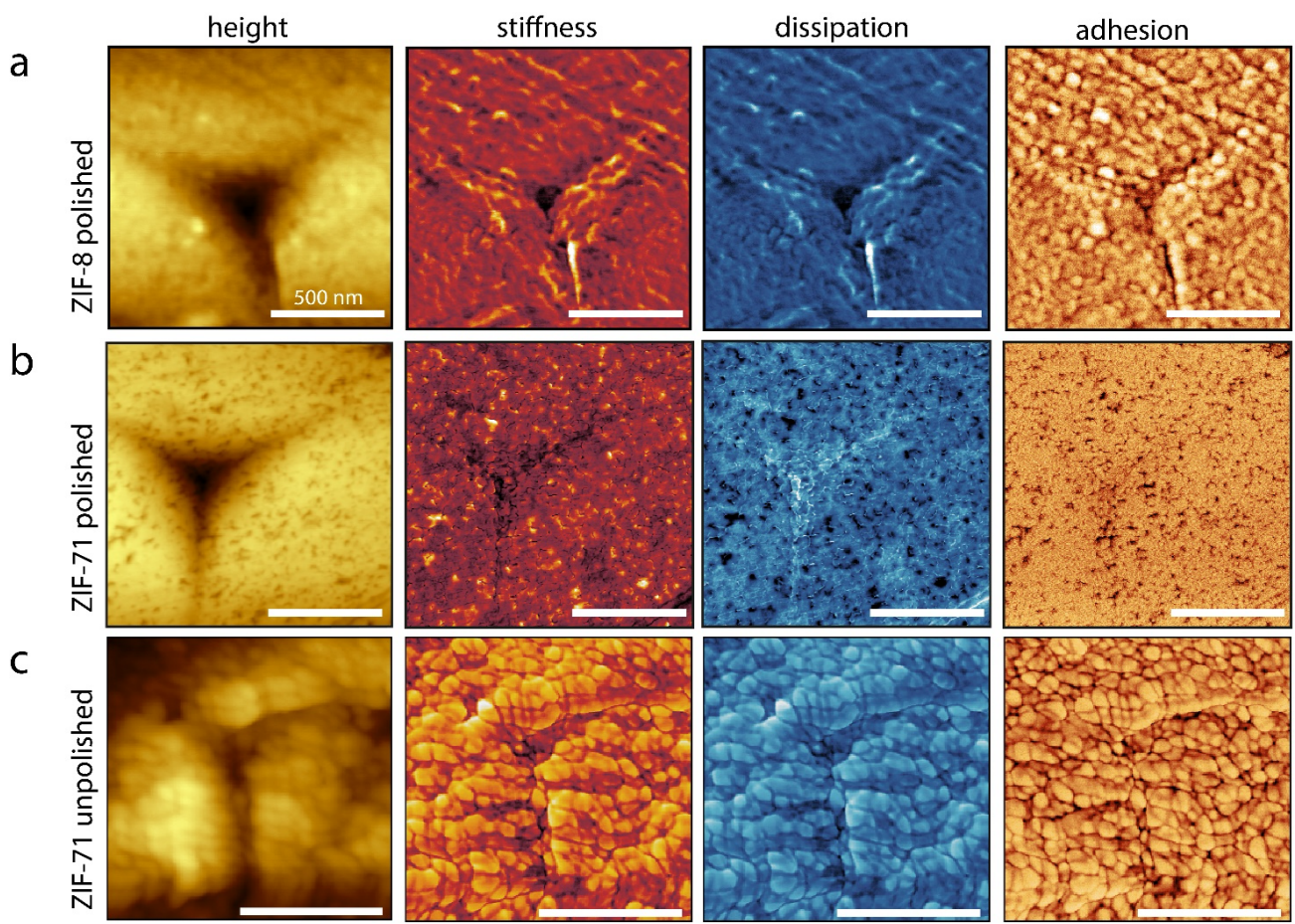

*Figure 4.* AFM height topography (left panels) and TFM results showing the stiffness, dissipation, and adhesion maps of residual indents on the surface of the ZIF-8 and ZIF-71 monoliths. The scale bar is 500 nm. The polished surface of ZIF-71 reveals prominent intergranular porosity.

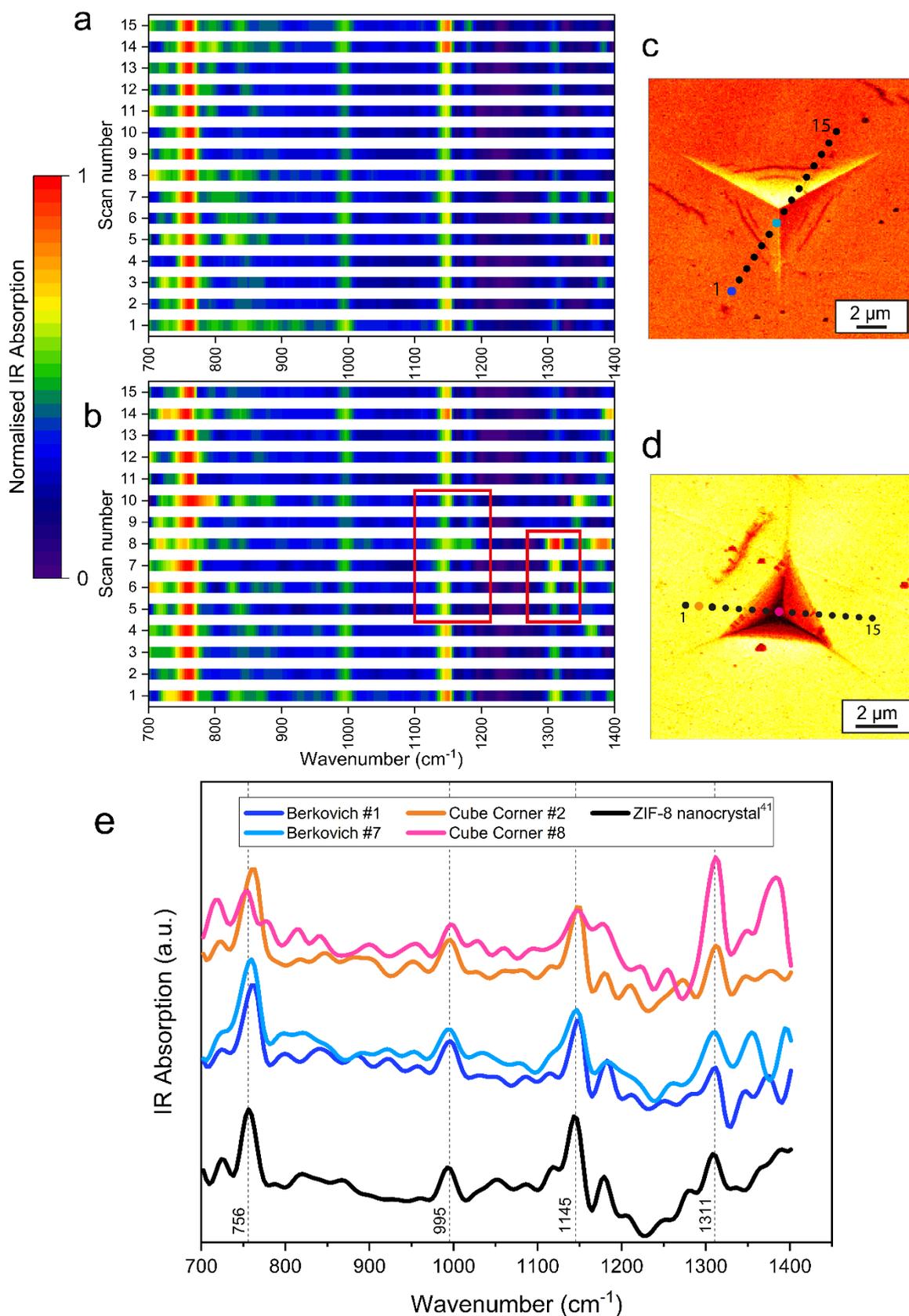

***Figure 5.*** *nanoFTIR absorption spectra of ZIF-8 monolith taken across (a) a Berkovich and (b) a cube corner residual indent, corresponding to the locations highlighted in (c) and (d), respectively. (e) Two spectra of each set, one measured outside and one inside the indent, were plotted and compared to the nanoFTIR spectra of an undeformed (pristine) ZIF-8 nanocrystal (adapted from [41]).*

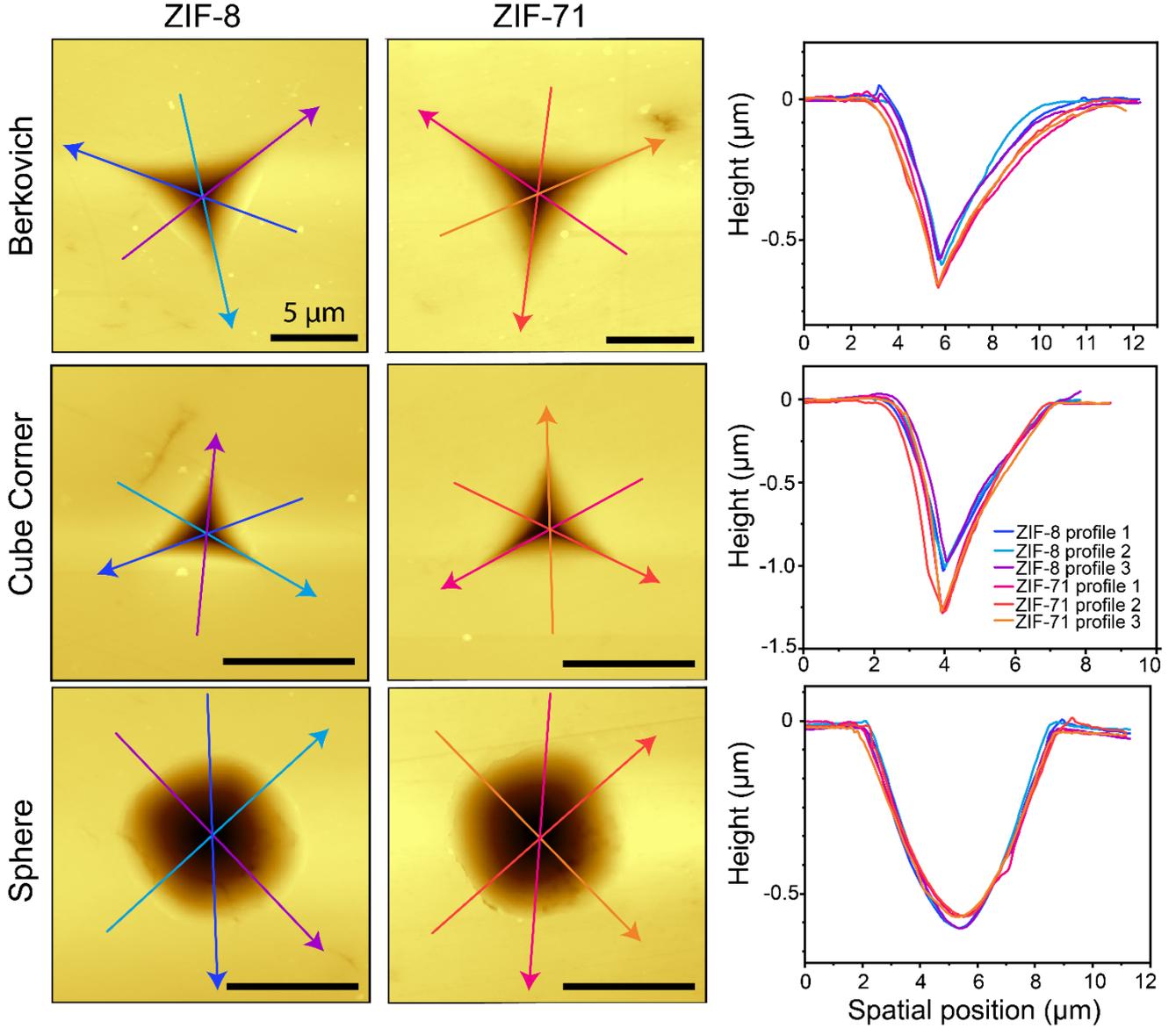

***Figure 6.*** *AFM height topography of the residual indents of Berkovich, cube corner, and spherical indentations for ZIF-8 and ZIF-71 monoliths. The corresponding cross-sectional profiles along the designated paths are plotted on the right panels. The scale bar is 5 μm.*

*Fracture toughness*

The fracture toughness ($K_{IC}$) of the ZIF-8 and ZIF-71 monoliths was evaluated by the indentation crack length method, that relies on nanoindentation measurements obtained by a cube corner indenter. Assuming a Palmqvist crack configuration, the Laugier's formula[42] was used:

$$K_{IC} = k \left(\frac{a}{l}\right)^{\frac{1}{2}} \left(\frac{E}{H}\right)^{\frac{2}{3}} \frac{P}{c^{\frac{3}{2}}} \quad (2)$$

where is $k$ is an empirical constant of the indenter shape ($k = 0.057$ for cube corner [43]), $a$ is the distance between the center and the tip of the indent, $l$ is the crack length starting from the corner of the indent, $E$ and $H$ are the Young's modulus and hardness respectively, $P$ is the maximum load, and $c = a + l$.

The obtained values of $K_{IC}$ are shown in *Table 2*. Between the two monoliths being studied here, we show the ZIF-8 monolith is relatively easier to crack: the radial cracks propagating from the indent's corners are visible already with a maximum depth of 2 μm (*Figure 7*). ZIF-71 turned out to be twice as tough as ZIF-8 and a maximum indentation depth of 5 μm was required in order to detect any cracks. The fracture toughness obtained are consistent with the few values available in literature, which reported the toughness of other classes of MOF materials, in the form of single crystals (0.08 to 0.33 MPa$\sqrt{m}$, for layered 2D to dense 3D frameworks),[10] and ZIF-62 glass by melting method (approximately 0.1 MPa$\sqrt{m}$ ).[44]

| Monolith | Fracture toughness $K_{IC}$ (MPa$\sqrt{m}$) |
|---|---|
| ZIF-8 | 0.074 ± 0.023 |
| ZIF-71 | 0.145 ± 0.050 |

**Table 2.** *Values of the fracture toughness $K_{IC}$ determined for the two monoliths using the residual indents generated by a cube corner indenter in Figure 7.*

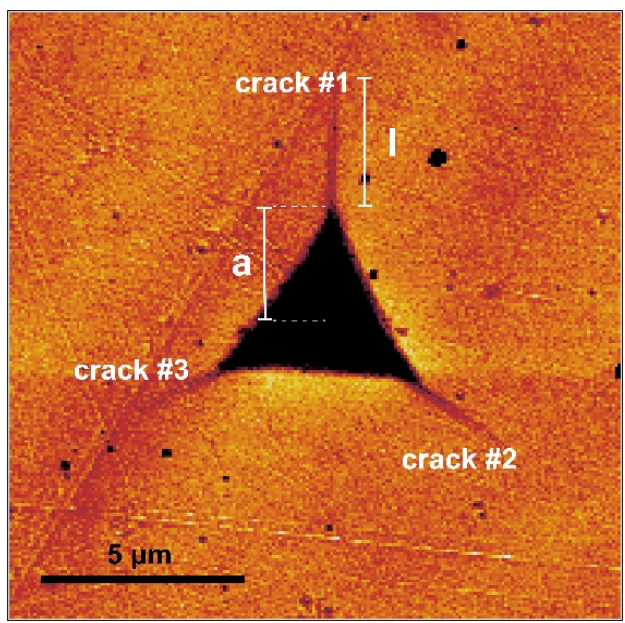

**Figure 7.** *Indentation-induced crack length measurement method for a cube corner residual indent on the surface of a ZIF-8 monolith.*

**Conclusions**

Two types of zeolitic MOF monoliths (ZIF-8 and ZIF-71) have been synthesized by a simple sol-gel process, a low-cost technique to produce MOF monoliths that does not require any support materials such as templates or binders and can be done at room temperature and without the need of high compacting pressures. The mechanical properties were studied by means of Berkovich, cube corner, and spherical nanoindentation, AFM and TFM imaging, where the experimental findings are further substantiated by FE simulations. The mechanical behavior of the material is well approximated by a simple elastic-perfectly plastic material, exhibiting a good ductility in compression. This result, together with the fine-grained nanostructure observed by TFM, suggests that the monoliths initially deform by grain boundaries sliding. When the stress increases, some densification occurs, due to the failure of the framework, presumably triggered by shear, in the contact area, as demonstrated by nanoFTIR measurements of the local vibrational spectra. Our findings provide a starting-point to gain new understanding of the mechanical behavior of sol-gel MOF monoliths, which is crucial for the transition of this class of materials from academia to practical applications.

**Methods**

*Materials*

Zinc nitrate hexahydrate ($Zn(NO_3)_2 \cdot 6H_2O$), zinc acetate dihydrate ($Zn(CH_3CO_2)_2 \cdot 2H_2O$), 2-methylimidazole (mIm), 4,5-dichloroimidazole (dcIm), triethylamine ($NEt_3$), dimethylformamide (DMF), methanol (MeOH), acetonitrile (MeCN) were purchased from Fisher Scientific and used as received.

*Synthesis of monoliths*

ZIF-8 ($Zn(mIm)_2$) monoliths were synthesised according to the following procedure. 0.595 g of $Zn(NO_3)_2 \cdot 6H_2O$ and 0.493 g of mIm were dissolved in 9 mL of DMF each and stirred for 5 minutes. Then, 0.837 mL of $NEt_3$ were added to the linker solution. Subsequently, the two solutions were mixed in a 50 mL vial, where a gel was promptly formed. The molar ratio $Zn(NO_3)_2 \cdot 6H_2O$ : mIm : DMF : $NEt_3$ is 1 : 3 : 116 : 3. The mixture was sonicated for 5 minutes and then washed three times, in 50 mL of solvent (DMF, MeOH and MeCN, respectively), by centrifugation at 8000 RPM. The collected solid was dried at room temperature (~25 °C) for 3 days under the fume cupboard.

ZIF-71 (Zn(dcIm)$_2$) monoliths were synthesized following the same procedures for ZIF-8 monoliths, starting from 0.439 g of Zn(CH$_3$CO$_2$)$_2$·2H$_2$O and 0.822 g of dcIm. The compound was washed three times in MeOH.

The PXRD patterns (Figures S1 and S2) confirm the successful synthesis of the frameworks, since all the main characteristic Bragg diffraction peaks of both ZIF-8 and ZIF-71 are present.

*Nanoindentation tests*

In a nutshell, a nanoindentation test is conducted by pressing a hard tip whose mechanical properties are known (usually diamond) into a sample whose properties are unknown. The result of the test is a load-depth curve, from which the Young's modulus (*E*) and hardness (*H*) can be determined. A load-depth curve consists of two segments: elasto-plastic loading and elastic unloading. If no plastic deformation occurs, the two segments overlap. Following the approach proposed by Oliver and Pharr,[36] modulus and hardness can be calculated using the following equations:

$$E^* = \frac{\sqrt{\pi}}{2} \frac{S}{\sqrt{A(h_{max})}} \qquad (3)$$

$$H = \frac{P_{max}}{A(h_{max})} \qquad (4)$$

where $E^*$ is the contact modulus, *S* is the contact stiffness (slope of the unloading curve at maximum load), *A(h)* is the area function, and $P_{max}$ and $h_{max}$ are the maximum load and depth, respectively. The contact modulus ($E^*$) is a function of the Young's moduli and Poisson's ratios of the sample ($E_s$, $v_s$) and the indenter ($E_i$, $v_i$):

$$\frac{1}{E^*} = \frac{1-v_s^2}{E_s} + \frac{1-v_i^2}{E_i} \qquad (5)$$

When the indenter is significantly stiffer than the sample, the second term on the right-hand side of *Equation 4* can be neglected. The area function *A(h)* is a 3$^{rd}$ order polynomial that relates the contact area to the contact depth, and it is determined through calibration using a fused silica sample.

The hardness is considered to be proportional to the yield stress ($\sigma_Y$):[45]

$$H = C \cdot \sigma_Y \qquad (6)$$

where *C* is called the "constraint factor" and is material dependent. It usually lies between 1.5 (in glasses, characterized by a small $E/\sigma_Y$ ratio) and 3 (in metals, large $E/\sigma_Y$ ratio).

An extension of the Oliver and Pharr method is represented by the Continuous Stiffness Measurements (CSM).[36, 46] This technique superimposes a 2-nm oscillation on the quasi-static force, using a frequency-specific amplifier to measure the response of the indenter. CSM enables the measurement of mechanical properties as a continuous function of the penetration depth into the surface.

The nanoindentation tests were conducted using an iMicro nanoindenter (KLA-Tencor). The as-synthetized monoliths were cold mounted in epoxy resin (Struers Epofix), resulting in a cylindrical sample, suitable for nanoindentation. In order to get reliable results from the indentation tests, the contact surface must be flat. Therefore, the specimen surface was thoroughly polished with sandpapers and diamond suspensions.

Berkovich, spherical and cube corner diamond indenter tips were used. The Berkovich tip is a three-sided pyramid, generally used for small-scale measurements of Young's modulus and hardness. The spherical tip provides a smooth transition from elastic to plastic contact, which makes it particularly suitable for probing soft materials. The spherical tip used for this study has a radius of 4.53 μm. Cube corner is a three-sided pyramid with mutually perpendicular faces. The sharpness of the cube corner produces much higher stresses and strains in the region of contact. This results in the formation of well-defined cracks around the indent in brittle materials, which can be used to measure fracture toughness at small scales.

*Density measurements*

The density of the monoliths was measured using a Mettler Toledo laboratory balance equipped with a Density Kit. By exploiting the Archimedes' principle, the density of the sample was derived by measuring its weight in air and in an auxiliary liquid (distilled water).

*AFM imaging*

The surface topography of indents was measured by atomic force microscopy (AFM) as implemented in a neaSNOM instrument (neaspec GmbH) under the tapping-mode. A Scout350 (NuNano) probe was employed, with a nominal tip radius of 5 nm and a resonant frequency of 350 kHz.

*TFM imaging*

Tip Force Microscopy (TFM) is a measuring mode for the AFM to measure nanomechanical properties of the sample along with the topography. The TFM mode extends the capabilities of AFM, allowing to image properties such as local stiffness and adhesion, resulting in high contrast images.[47, 48] Height topography and stiffness, dissipation and adhesion maps were taken using the neaSNOM instrument (neaspec GmbH), set to Tip Force mode. An Arrow-NCR probe (NanoWorld) was employed, with a nominal tip radius of < 10 nm and a resonance frequency of 285 kHz.

*nanoFTIR*

Nano-Fourier-transform infrared (nano-FTIR) spectroscopy was performed using a neaSNOM instrument (Neaspec GmbH), where a platinum-coated AFM probe (Arrow-NCPt, tip radius < 25 nm, 285 kHz) under the tapping mode is illuminated by a broadband mid-IR laser source (Toptica). A line scan of 15-point spectra was acquired along the residual indents of the Berkovich and cube corner indentations on the ZIF-8 monolith. Each point spectra was acquired by Fourier-transform spectroscopy as an average of 11 individual interferograms taken on the same spot (spot size 20 nm), with 1024 pixels and an integration time of 12 ms per pixel, normalized by a reference spectrum taken on a silicon wafer.

*Finite element (FE) modelling*

For the finite element simulation of the nanoindentation test, a 2D axisymmetric model in Abaqus/Standard[49] was employed. The Berkovich and cube corner indenters can be modelled as conical indenters with semi-apical angles of 70.3° and 42.3° respectively, which result in the same area function of their pyramidal counterparts. As shown by Lichinchi *et al.*,[50] the difference between the load-depth curves obtained from the 2D axisymmetric and the 3D Berkovich model is negligible. A comparison between the 2D and 3D load-depth curves resulting from a simulated Berkovich indentation of a ZIF-8 monolith is shown in Figure S3. Considering that the 2D simulation is computationally less expensive, this is highly effective for simulating nanoindentation tests with a Berkovich indenter. All indenter geometries (including spherical) were modelled as rigid bodies. This approximation is justified since the real indenter, made of diamond, has a Young's modulus about 1000 times larger than the tested material.

The sample was meshed with 12692 CAX4 type quadrilateral elements, adapting the technique proposed in the *Abapy* package in *Python*.[51] A finer mesh was used in the contact area, in order to obtain an accurate representation of the stress distribution under the indenter tip. A coarser mesh was used further away from the contact zone to reduce the computational time (*Figure 1a*).

The following boundary conditions were imposed: the nodes along the axis of rotation are free to move only along such an axis while for the nodes at the bottom of the sample all the degrees of freedom were set to zero ("encastre" boundary condition).

The indenter-sample interaction was modelled with a "surface-to-surface" contact discretization. The contact constraint is imposed by defining the "master" and "slave" surfaces. The slave surface cannot penetrate the master surface and the direction of contact is perpendicular to the master surface. The slave surface is usually the softer of the two and needs to be meshed with finer elements.[49] We chose the sample surface as the slave surface.

The nanoindentation test was simulated by using two subsequent load steps, one for the loading part and the other for the unloading part. During the loading step, the indenter tip moves down along the axis of symmetry until the maximum depth (2000 nm) is reached. The return of the tip to its original position takes place during the unloading step.

The mechanical behavior of the samples was simulated by two different models: elastic-perfectly plastic and Drucker-Prager. The first one requires only three parameters to fully describe the elasto-plastic behavior of the material (Poisson's ratio, Young's modulus and yield stress). The second one is used for pressure-dependent materials, whose yield behavior depends on the hydrostatic pressure. This model requires three more material parameters, namely "angle of friction", "flow stress ratio" and "dilation angle",[49] that were set to 10, 1, and 10, respectively.

## Acknowledgements

We thank the ERC Consolidator Grant (PROMOFS grant agreement 771575) for funding this research.

## Supporting Information

X-ray diffraction data, comparison of 2D vs. 3D finite element simulations of the nanoindentation test, selection of optimal constraint factor from finite element simulated load-depth curves, shear stress contours of ZIF-71 finite element models